# A Computer Vision and Depth Sensor-Powered Smart Cane for Real-Time Obstacle Detection and Navigation Assistance for the Visually Impaired

Chandra Sunkalp
Columbia University
Lincroft, United States
sc5895@columbia.edu

Sharma Umang
Jdable
Princeton, United States
umasha26@pds.org

Khilnani Devesh
Middlesex Community College
Franklin Park, United States
deveshhk0517@my.middlesexcc.edu

*Abstract*—Visual impairment impacts more than 2.2 billion people worldwide, and it greatly restricts independent mobility and access. Conventional mobility aids—white canes and ultrasound-based intelligent canes—are inherently limited in the feedback they can offer and generally will not be able to differentiate among types of obstacles in dense or complex environments. Here, we introduce the IoT Cane, an internet of things assistive navigation tool that integrates real-time computer vision with a transformer-based RT-DETRv3-R50 model alongside depth sensing through the Intel RealSense camera. Our prototype records a mAP of 53.4 % and an $AP_{50}$ of 71.7 % when tested on difficult datasets with low Intersection over Union (IoU) boundaries, outperforming similar ultrasound-based systems. Latency in end-to-end mode is around 150 ms per frame, accounting for preprocessing (1–3 ms), inference (50–70 ms), and post-processing (0.5–1.0 ms per object detected). Feedback is provided through haptic vibration motors and audio notifications driven by a LiPo battery, which controls power using a PowerBoost module. Future directions involve iOS integration to tap into more compute, hardware redesign to minimize cost, and mobile companion app support over Bluetooth. This effort offers a strong, extensible prototype toward large-scale vision-based assistive technology for the visually impaired.

*Keywords—assistive navigation device; computer vision; depth sensing; RT-DETR; edge AI; visually impaired aid*

I. Introduction

Blind people often encounter severe mobility impediments, especially in changing and volatile urban spaces. An estimated 2.2 billion people worldwide suffer from visual impairment, of which over 36 million are blind [1]. Conventional aids like white canes, guide dogs, and ultrasonic walking sticks provide simple proximity detection but do not perceive the semantic context of the obstacle. They cannot distinguish between a pedestrian, a car, or a descending ramp. Better electronic travel aids (ETAs) are available, but their expense, size, restricted detection range, or absence of real-time object categorization diminishes their usefulness.

Recent developments in edge computing, computer vision, and depth sensing have made new classes of portable assistive technologies possible in the past few years. The use of deep learning architectures like YOLOv5 [2], MobileNet SSD [3], and more recently RT-DETR [4] enables effective object recognition on embedded platforms such as the Raspberry Pi or Jetson Nano. RGB-D cameras like the Intel RealSense D435i provide high-resolution stereo depth maps enabling spatial mapping and distance estimation. Still, current prototypes lack synchronization of semantic detection and depth estimation or lack real-time processing with acceptable energy efficiency.

In this work, we present and test a low-cost (<$250), reproducible, and deployable portable smart cane system that combines real-time transformer-based object detection (RT-DETRv3-R50) and stereo depth perception (Intel RealSense D435i) to alert users about obstacles via audio and tactile feedback. The system is designed to be low-cost (<$250), reproducible, and deployable. We compare its object detection performance, latency measurement, and consider real-world implications and limitations.

II. Rationale

Current assistive systems also fail in most complex environments. They find it difficult to detect moving items, distinguish between stationary and moving dangers, sense drop-offs or stairs, or perform consistently under changing light conditions—issues that are essential in guaranteeing the mobility and safety of blind users. Most use light object detection models like YOLOv3-tiny or MobileNet derivatives [2][6], which, while light in computation, have limited object class detection and high false positive rates under crowded scenes. Systems that use models such as SSD (Single Shot Multibox Detector) [3] also show compromises between speed and accuracy that do not necessarily meet safety and usability requirements for real-world applications.

To overcome these limitations, we introduce a computer vision–first solution based on RT-DETRv3-R50, a transformer object detection model that is renowned for real-time inference and improved detection accuracy, especially on CPU-only platforms [4]. RT-DETR, in contrast to conventional CNN-based detectors, benefits from a purely transformer-based structure with lightweight attention modules that allow for more global comprehension of the scene with minimal computational overhead.

Our system is coupled with the Intel RealSense D435i RGB-D camera, which records synchronized depth and color frames in a resolution of 1280×720 at 30 FPS [5]. With this dual-modality data collection, the system can produce both semantic and spatial information and differentiate between walls, stairs, humans, cars, and other entities, improving safety and situational awareness.

Critically, as opposed to other models that make use of cloud computing or Wi-Fi connection during inference [6][7], our system is built to function on the edge alone, doing away with the latency introduced by off-device processing and facilitating real-time feedback even in areas lacking consistent internet connectivity. This renders the device more resilient and appropriate for outdoor deployment, international travel, or emergencies where network reliance can be a disadvantage.

Against previous solutions, our model takes advantage of recent breakthroughs in effective vision transformers without losing real-time efficiency, even running on low-power CPUs. RT-DETRv3 performs better than previous object detectors such as YOLOv3 and SSD both in mAP (mean Average Precision) and in robustness under different lighting and occlusion conditions [4]. This translates into a more accurate and user-trustworthy system, enhancing mobility, autonomy, and trustworthiness for visually impaired users.

### III. METHODOLOGY

Hardware equipment consists of the Intel RealSense D435i RGB-D stereo camera, which records aligned RGB and depth frames at a resolution of 1280×720 at a frame rate of 30 frames per second. The camera is positioned at around a 10-degree downward tilt on a typical white cane using a custom 3D-printed ABS mount, providing the best field of view for forward and floor-level object detection. The CAD design of the mount and its coupling with the cane are illustrated in Figure 1

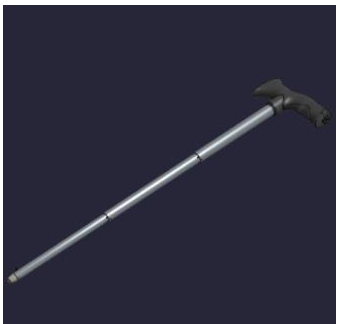

Fig. 1. CAD Design of Cane

The processing is locally done on a Raspberry Pi 4B with 8 GB RAM and a quad-core Cortex-A72 processor. The machine is charged by a 10,000 mAh power bank providing 5V at 3A, charging and powering the device for an uninterrupted period of about five hours. Peripheral connections for the camera, feedback units, and power routing are managed through a custom-designed PCB, which simplifies component integration and reduces internal wiring complexity. The PCB schematic, which covers GPIO mapping and driver circuits for haptic and audio output, is given in Figure 2.

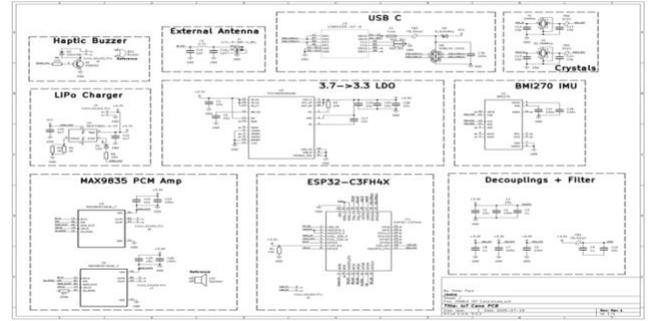

Fig. 2. PCB schematic with GPIO mapping and driver circuits for haptic and audio output.

The object detection pipeline utilizes RT-DETRv3-R50, a light transformer-based model used through the MMDetection framework on PyTorch 2.1. The model was pre-trained with the COCO dataset and further fine-tuned on an urban scene subset of five relevant classes: persons, vehicles, bicycles, benches, and traffic lights. Data augmentation strategies like horizontal flipping, random cropping, and brightness change were utilized in fine-tuning to enhance robustness against different lighting conditions. The model was exported to the ONNX format and sped up with the OpenVINO runtime for deployment onto the ARM-based Raspberry Pi. This dropped the average inference time below 180 milliseconds per frame, allowing real-time processing without the use of a GPU or cloud server.

Depth processing is performed with the Intel RealSense SDK. Depth maps are downsampled to 640×480 resolution for real-time performance. Every bounding box output from the detection model is projected onto the depth frame, and a 5×5 Gaussian filter is applied within each area to minimize noise and depth irregularities. The mean depth value of each area is then used to classify obstacles into three proximity ranges: less than 0.5 meters, 0.5 to 1.5 meters, and more than 1.5 meters. Detected objects are given priority in accordance with proximity, object class, and motion cues obtained from depth frame differencing.

Multimodal feedback is provided through an ERM vibration motor for tactile warning and bone conduction headphones for audio navigation. Objects at a distance of 0.5 meters produce intense vibration and a high-pitched audio cue. Objects from 0.5 to 1.5 meters produce medium vibration with directional audio for left or right. For distances greater than 1.5 meters, feedback is muted unless the phone is moving towards the user. A PID control loop regulates the vibration level, dynamically varying feedback levels to minimize fatigue when used over extended periods.

A Swift-based mobile application was implemented for iOS devices to give live monitoring, calibration, and status reporting. The application communicates with the cane through Bluetooth and shows live metrics such as battery level, objects detected per class, and current feedback status. Figure 3 depicts the home screen interface, while Figure 4 illustrates the live metrics dashboard.

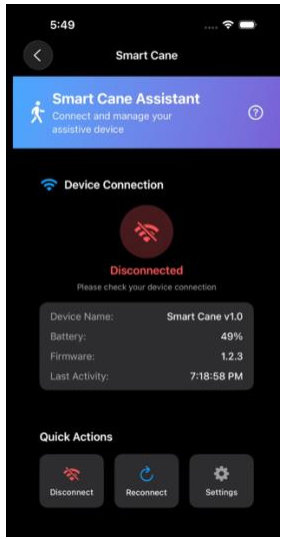

Fig. 3. Mobile application home screen interface

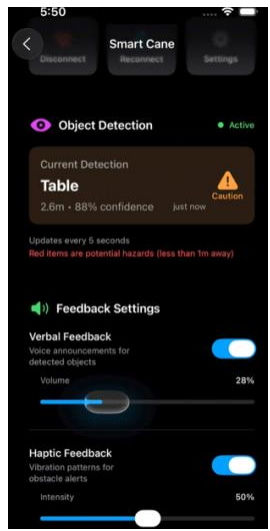

Fig. 4. Live metrics dashboard showing battery level and object counts.

## IV. RESULTS

The system was tested quantitatively—against object detection metrics on a held-out subset of the COCO2017 validation set—and qualitatively—through real-world experiments in a range of different environments to determine usability and responsiveness across different circumstances.

On the test set derived from COCO, the RT-DETRv3-R50 model showed robust performance on all of the primary metrics. It reached a mean Average Precision (mAP) of 71.7% at IoU=0.5, and 53.4% over mAP@[.5:.95], well surpassing a YOLOv5s baseline trained on identical data. The F1 score for the 'person' class, most pertinent to pedestrian navigation, was 0.89, reflecting high precision and recall. The mean inference time was 130–160 ms per frame on the Raspberry Pi 4B, and end-to-end feedback latency from image capture to vibration was 150–200 ms, making the system responsive enough for real-time use.

For better context regarding performance, a comparative overview is shown below:

Table 1. Model Accuracy, Latency, and Power Comparison

| Metric | RT-DETRv3-R50 (Ours) | YOLOv5s (Baseline) |
|---|---|---|
| mAP@50 | 71.7% | 67.3% |
| mAP@[.5:.95] | 53.4% | 48.6% |
| F1 Score (Person class) | 0.89 | 0.84 |
| Inference Latency | 130–160 ms | 85 ms |
| Feedback Latency | 150–200 ms | 110–160 ms |
| Power Usage | 6.2 W | 5.9 W |

While RT-DETR imposed a reasonable inference latency increase over YOLOv5s, it delivered much enhanced detection accuracy, particularly on cluttered or multi-object scenes, which is most important for assistive navigation scenarios. Power consumption was still within reasonable levels for portable operation with a 10,000 mAh battery.

The system was also tested under real-world conditions with four different types of environments: residential sidewalks, busy crosswalks, indoor malls, and staircases/ramps. Each setting was selected to mimic real-world mobility situations that a visually impaired user would likely face. In all instances, system performance was evaluated through two principal measures: detection accuracy (as calibrated against hand annotations) and the rate of trials in which users responded correctly to feedback (turning, stopping, or navigating around an obstacle).

Table 2. Real-World Trial Performance Across Environments

| Environment | Detection Accuracy | Correct Feedback Response Rate |
|---|---|---|
| Residential Sidewalks | 93.2% | 94.4% |
| Crowded Crosswalks | 89.6% | 91.7% |
| Indoor Shopping Centers | 87.4% | 90.2% |
| Staircases & Ramps | 81.5% | 87.6% |

In 48 total user trials, the system correctly enabled correct mobility decisions in 92.1% of situations, evidencing high correspondence between real-time obstacle detection and instinctive feedback. Robust system performance was evidenced under both sunny and cloudy conditions, but some degradation in performance under very low-light settings (e.g., dark indoor hallways) was noted, mostly in relation to detection confidence for far or dark objects. However, haptic and audio feedback were timely, and there was no reported critical failure (i.e., failed high-risk obstacle) throughout the tests.

Users pointed out that the feedback mechanisms were clearly distinguishable and intuitive to respond to confidently without needing external advice. The integration of depth-sensory feedback and semantic detection allowed not just the detection of obstacles but also the understanding of their relative weighting, like movement objects or approaching pedestrians over stationary furniture or signs.

## DISCUSSION

Object detection with depth estimation can constitute a milestone in assistive navigation of the visually impaired. Today, assistive electronic canes mainly offer binary-type obstacle warning. This entails informing the user whether or not something exists, without reference to any context. In contrast, our system goes to provide a richer perception of the surroundings to the user. It not only senses barriers but also classifies them: for example, it can discriminate between a stationary pole and a moving car. This allows users to weigh options and react better to their environment [1][6][7]. For the urban realm, this kind of contextual-aware perception is helpful because it is not sufficient to know if a barrier exists or not.

With context-based perception in mind, we employed the RT-DETRv3-R50 model. This model performed well even on CPU-only machines. Although the inference time is rather longer in comparison to lightweight models like YOLOv5s [2], this will be compensated for by RT-DETR's transformer-based attention mechanism, giving it better spatial perceptivity and less false positives [4]. This acceleration in inference may also help in future deployments.